\documentclass[final]{aipprocm}
\usepackage{graphicx}
\layoutstyle{8x11single}

\newcommand{\ind}[2]{^{\mbox{\tiny $#1$}}_{\mbox{\tiny #2}}}
\newcommand{\DP}{\Delta\Pi}
\newcommand{\Nc}{N_{\mbox{\tiny c}}}
\newcommand{\nf}{n_{\mbox{\tiny f}}}
\newcommand{\nfs}{n_{\mbox{\tiny f}}}
\newcommand{\DeltaQCD}[1]{\Delta^{\mbox{\tiny #1}}_{\mbox{\tiny QCD}}}
\newcommand{\RTau}[1]{R_{\tau, \mbox{\tiny #1}}}
\newcommand{\Vud}{V_{\mbox{\tiny ud}}}
\newcommand{\Sew}{S_{\!\mbox{\tiny EW}}}
\newcommand{\dpew}{\delta'_{\mbox{\tiny EW}}}
\newcommand{\va}{_{\mbox{\tiny V/A}}}
\newcommand{\cva}{\chi\va}
\newcommand{\MTau}{M_{\tau}}

\begin{document}

\title{Inclusive $\tau$~lepton hadronic decay in vector and
axial--vector channels within dispersive approach to~QCD}



\author{A.V.~Nesterenko}{address={Bogoliubov Laboratory of Theoretical Physics,
Joint Institute for Nuclear Research, \\
Dubna, 141980, Russian Federation \\
E-mail: nesterav@theor.jinr.ru}}

\begin{abstract}
The dispersive approach to QCD, which properly embodies the intrinsically
nonperturbative constraints originating in the kinematic restrictions on
relevant physical processes and extends the applicability range of perturbation
theory towards the infrared domain, is briefly overviewed. The study of OPAL
(update~2012) and ALEPH (update~2014) experimental data on inclusive
$\tau$~lepton hadronic decay in vector and axial--vector channels within
dispersive approach is presented.
\end{abstract}

\maketitle

The theoretical particle physics widely employs the methods based on dispersion
relations. In particular, such methods have proved to be efficient in the
extension of the range of applicability of chiral perturbation
theory~\cite{Portoles, Passemar}, assessment of the hadronic light--by--light
scattering~\cite{DispHlbl}, precise determination of parameters of
resonances~\cite{Kaminski}, and many other issues.

The dispersion relations render the kinematic restrictions on pertinent
physical processes into the mathematical form and impose stringent
nonperturbative constraints on relevant quantities, such as the hadronic
vacuum polarization function~$\Pi(q^2)$. These constraints have been
properly embodied within dispersive approach to QCD\footnote{Its
preliminary formulation was discussed in Refs.~\cite{DQCDPrelim1,
DQCDPrelim2}.}~\cite{DQCD1a, PRD88}, which provides unified integral
representations for~$\Pi(q^2)$, related function~$R(s)$, which is
identified with the so--called \mbox{$R$--ratio} of electron--positron
annihilation into hadrons, and Adler function~$D(Q^2)$:
\begin{eqnarray}
\label{P_DQCD}
\DP(q^2,\, q_0^2) &=& \DP^{(0)}(q^2,\, q_0^2) +
\!\int_{m^2}^{\infty} \rho(\sigma)
\ln\biggl(\frac{\sigma-q^2}{\sigma-q_0^2}
\frac{m^2-q_0^2}{m^2-q^2}\biggr)\frac{d\,\sigma}{\sigma},\quad \\[1mm]
\label{R_DQCD}
R(s) &=& R^{(0)}(s) + \theta(s-m^2) \int_{s}^{\infty}\!
\rho(\sigma) \frac{d\,\sigma}{\sigma}, \\[1mm]
\label{Adler_DQCD}
D(Q^2) &=& D^{(0)}(Q^2) +
\frac{Q^2}{Q^2+m^2} \int_{m^2}^{\infty} \rho(\sigma)
\frac{\sigma-m^2}{\sigma+Q^2} \frac{d\,\sigma}{\sigma}.
\end{eqnarray}
In these equations $m$~denotes the value of hadronic production threshold,
$\rho(\sigma)$ is the spectral density
\begin{equation}
\label{RhoGen}
\rho(\sigma) = \frac{1}{2 \pi i} \frac{d}{d\,\ln\sigma}
\lim_{\varepsilon \to 0_{+}}
\Bigl[p(\sigma-i\varepsilon)-p(\sigma+i\varepsilon) \Bigr]
= - \frac{d\,r(\sigma)}{d\,\ln\sigma}
= \frac{1}{2 \pi i} \lim_{\varepsilon \to 0_{+}}
\Bigl[d(-\sigma-i\varepsilon)-d(-\sigma+i\varepsilon) \Bigr]\!,
\end{equation}
$\DP(q^2\!,\, q_0^2) = \Pi(q^2) - \Pi(q_0^2)$ stands for the subtracted
hadronic vacuum polarization function, whereas $p(q^2)$, $r(s)$, and~$d(Q^2)$
denote the strong corrections to the functions~$\Pi(q^2)$, $R(s)$,
and~$D(Q^2)$, respectively. The derivation of integral representations
(\ref{P_DQCD})--(\ref{Adler_DQCD}) employs only the kinematic restrictions on
the relevant physical processes, the asymptotic ultraviolet behavior of the
hadronic vacuum polarization function, and requires neither additional
approximations nor phenomenological assumptions, see Refs.~\cite{DQCD1a,
PRD88}.

The common prefactor $\Nc\sum_{f=1}^{\nfs} Q_{f}^{2}$ is omitted throughout the
paper, where $\Nc=3$ is the number of colors, $Q_{f}$~stands for the electric
charge of $f$--th quark, and $\nf$~is the number of active flavors. In
Eqs.~(\ref{P_DQCD})--(\ref{Adler_DQCD}) $Q^2 = -q^2 > 0$ and $s = q^2 > 0$
denote the spacelike and timelike kinematic variables, respectively, and
$\theta(x)$ is the unit step--function [$\theta(x)=1$ if $x \ge 0$ and
$\theta(x)=0$ otherwise]. The leading--order terms in
Eqs.~(\ref{P_DQCD})--(\ref{Adler_DQCD}) read
\begin{eqnarray}
\label{P0L}
\DP^{(0)}(q^2,\, q_0^2) &=& 2\,\frac{\varphi - \tan\varphi}{\tan^3\varphi}
- 2\,\frac{\varphi_{0} - \tan\varphi_{0}}{\tan^3\varphi_{0}}, \\
\label{R0L}
R^{(0)}(s) &=& \theta(s - m^2)\biggl(1-\frac{m^2}{s}\biggr)^{\!\!3/2}, \\
\label{D0L}
D^{(0)}(Q^2) &=& 1 + \frac{3}{\xi}\Bigl[1 \!-\! \sqrt{1\!+\!\xi^{-1}}\,
\sinh^{-1}\!\bigl(\xi^{1/2}\bigr)\!\Bigr]\!,~ \qquad
\end{eqnarray}
where $\sin^2\!\varphi = q^2/m^2$, $\sin^2\!\varphi_{0} = q^{2}_{0}/m^2$,
and $\xi=Q^2/m^2$, see papers~\cite{PRD88, DQCD3, C12} and references
therein for the details.

There is still no unambiguous method to restore the complete expression for the
spectral density~$\rho(\sigma)$~(\ref{RhoGen}) (discussion of this issue can be
found in, e.g., Refs.~\cite{DQCD3, C12, PRD62}). Nonetheless, the perturbative
contribution to~$\rho(\sigma)$ can be calculated by making use of the
perturbative expression for either of the strong corrections to the functions
on hand (see, e.g., Refs.~\cite{CPC, BCmath}):
\begin{eqnarray}
\label{RhoPert}
\rho\ind{}{pert}(\sigma) = \frac{1}{\pi} \frac{d}{d\,\ln\sigma}\,
\mbox{Im}\lim_{\varepsilon \to 0_{+}} p\ind{}{pert}(\sigma-i\varepsilon)
= \! - \frac{d\, r\ind{}{pert}(\sigma)}{d\,\ln\sigma}
\!=\! \frac{1}{\pi}\, \mbox{Im}\lim_{\varepsilon \to 0_{+}}
d\ind{}{pert}(-\sigma-i\varepsilon).
\end{eqnarray}
In this paper the model~\cite{PRD88} for the spectral density will be
employed:
\begin{equation}
\label{RhoMod}
\rho(\sigma) = \frac{4}{\beta_{0}}\frac{1}{\ln^{2}(\sigma/\Lambda^2)+\pi^2} +
\frac{\Lambda^2}{\sigma},
\end{equation}
where $\beta_{0} = 11 - 2\nf/3$ and $\Lambda$ denotes the QCD scale parameter.
The first term on the right--hand side of Eq.~(\ref{RhoMod}) is the one--loop
perturbative contribution, whereas the second term represents intrinsically
nonperturbative part of the spectral density, see paper~\cite{PRD88} and
references therein for the details.

\begin{figure}[t]
\centerline{\includegraphics[width=80mm]{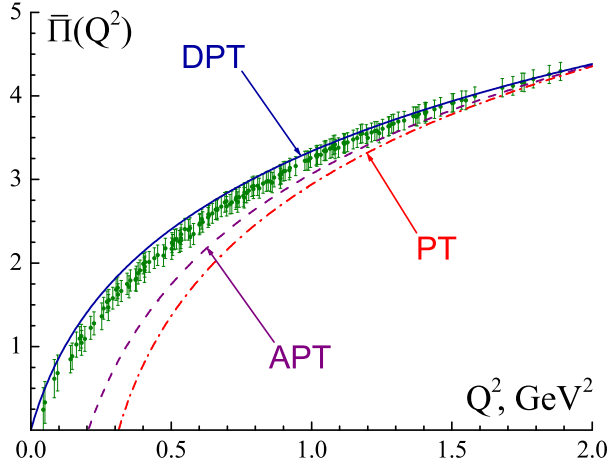}}
\caption{Comparison of the hadronic vacuum polarization
function~${\bar\Pi}(Q^2) = \DP(0,-Q^2)$ with relevant lattice simulation
data~\cite{Lat5}, see Ref.~\cite{DQCDEW} for the details.}
\label{Plot:PDQCD}
\end{figure}

It is worthwhile to mention also that in the massless limit ($m=0$) for the
case of perturbative spectral function [$\rho(\sigma) = \mbox{Im}\;
d\ind{}{pert}(-\sigma - i\,0_{+})/\pi$] two equations~(\ref{R_DQCD})
and~(\ref{Adler_DQCD}) become identical to those of the analytic perturbation
theory~(APT)~\cite{APT} (see also Refs.~\cite{APT1, APT2, APT3, APT4, APT5,
APT6, APT7a, APT7b, APT8, APT9, APT10, APT11}). However, it is essential to
keep the value of hadronic production threshold nonvanishing, since the
massless limit loses some of the substantial nonperturbative constraints, which
relevant dispersion relations impose on the functions on hand, see
Refs.~\cite{DQCD1a, PRD88, DQCDEW, DQCD1b}.

The dispersively improved perturbation theory~(DPT)~\cite{DQCD1a, PRD88}
extends the applicability range of perturbative approach towards the infrared
domain. In particular, the Adler function\footnote{The studies of Adler
function within other approaches can be found in Refs.~\cite{Maxwell, PeRa,
Kataev, MSS, Cvetic, BJ, Fischer1, Fischer2}.}~(\ref{Adler_DQCD}) conforms with
relevant experimental prediction in the entire energy range~\cite{DQCD1a,
DQCD1b, DQCD2} and the hadronic vacuum polarization function~(\ref{P_DQCD})
agrees with pertinent lattice simulation data~\cite{DQCDEW}, see
Fig.~\ref{Plot:PDQCD}. Furthermore, the
representations~(\ref{P_DQCD})--(\ref{Adler_DQCD}) conform with the results
obtained in Ref.~\cite{PRL99PRD77} as well as in Ref.~\cite{RCTaylor}.
Additionally, the respective hadronic contributions to the muon anomalous
magnetic moment and to the shift of the electromagnetic fine structure constant
at the scale of $Z$~boson mass evaluated in the framework of~DPT proved to be
in a good agreement with recent estimations of these quantities~\cite{DQCDEW}.
All this testifies to the efficiency of dispersive approach~\cite{DQCD1a,
PRD88} in the studies of nonperturbative aspects of the strong interaction.

\smallskip

The study of the inclusive $\tau$~lepton hadronic decay represents a
particular interest, since this process probes the \mbox{low--energy}
hadron dynamics. Specifically, the theoretical expression for the relevant
experimentally measurable quantity reads
\begin{equation}
\label{RTauGen}
\RTau{V/A}^{\tiny{J=1}} = \frac{\Nc}{2}\, |\Vud|^2\,\Sew\,
\Bigl(\DeltaQCD{V/A} + \dpew \Bigr).
\end{equation}
In this equation $|\Vud| = 0.97425 \pm 0.00022$ is
Cabibbo--Kobayashi--Maskawa matrix element~\cite{PDG2012}, $\dpew =
0.0010$ and $\Sew = 1.0194 \pm 0.0050$ denote the electroweak
corrections~\cite{EWF}, and
\begin{equation}
\label{DeltaQCDDef}
\DeltaQCD{V/A} = \frac{2}{\pi}\int_{m\va^2}^{\MTau^2}\!
\biggl(1-\frac{s}{\MTau^2}\biggr)^{\!\!2}\biggl(1+2\frac{s}{\MTau^2}\biggr)\,
\mbox{Im}\,\Pi^{\mbox{\tiny V/A}}(s+i0_{+})\, \frac{d s}{\MTau^2}
\end{equation}
stands for the hadronic contribution, see Refs.~\cite{BNPPDP, Tau}.
In~Eq.~(\ref{DeltaQCDDef}) $\MTau \simeq 1.777\,$GeV~\cite{PDG2012} is the mass
of $\tau$~lepton, whereas $m\va$~denotes the total mass of the lightest allowed
hadronic decay mode of $\tau$~lepton in the corresponding channel.

\begin{figure}[t]
\begin{tabular}{cc}
\includegraphics[width=72.5mm]{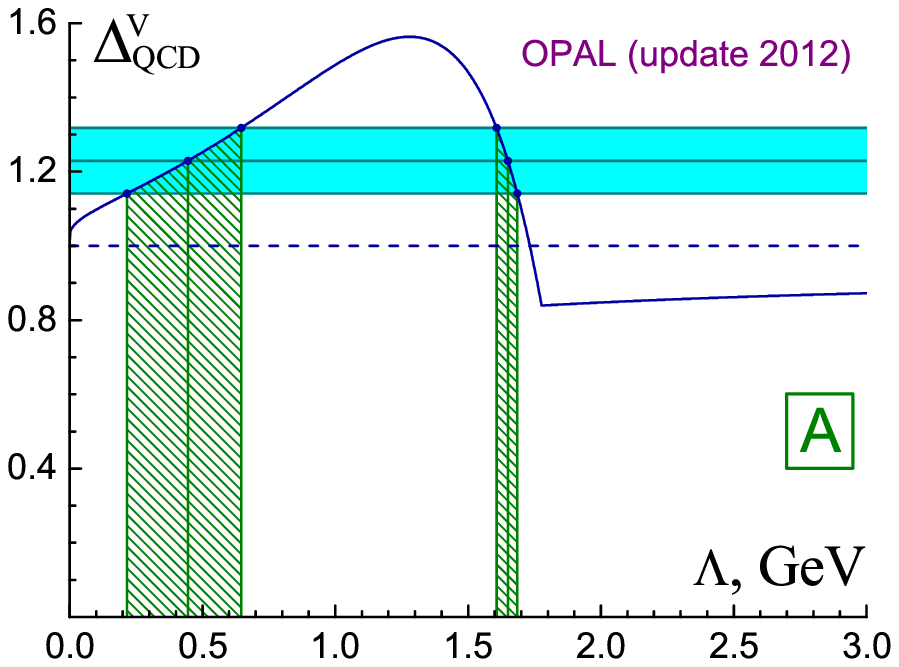}~ &
~\includegraphics[width=72.5mm]{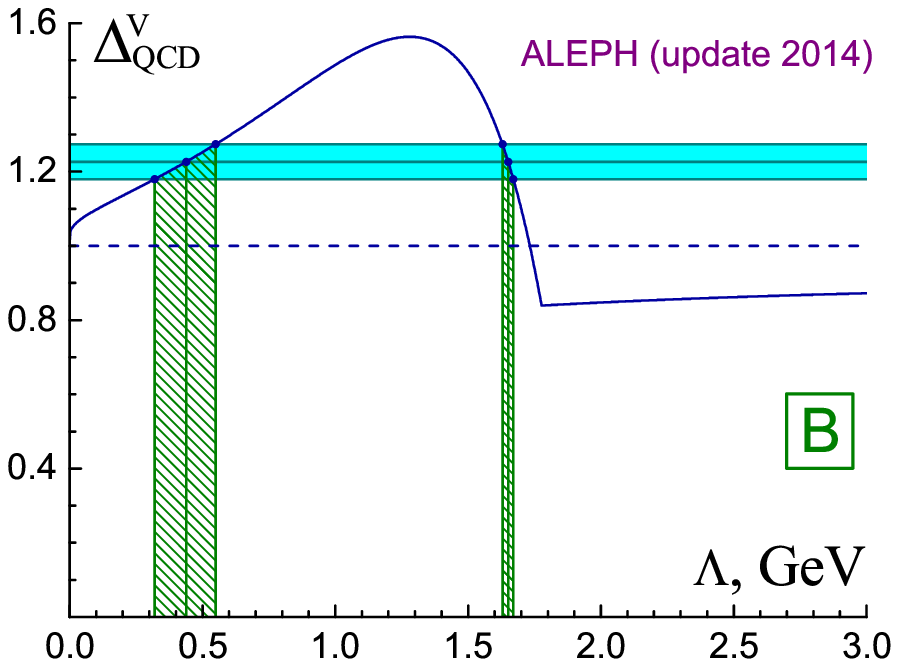} \\[5mm]
\includegraphics[width=72.5mm]{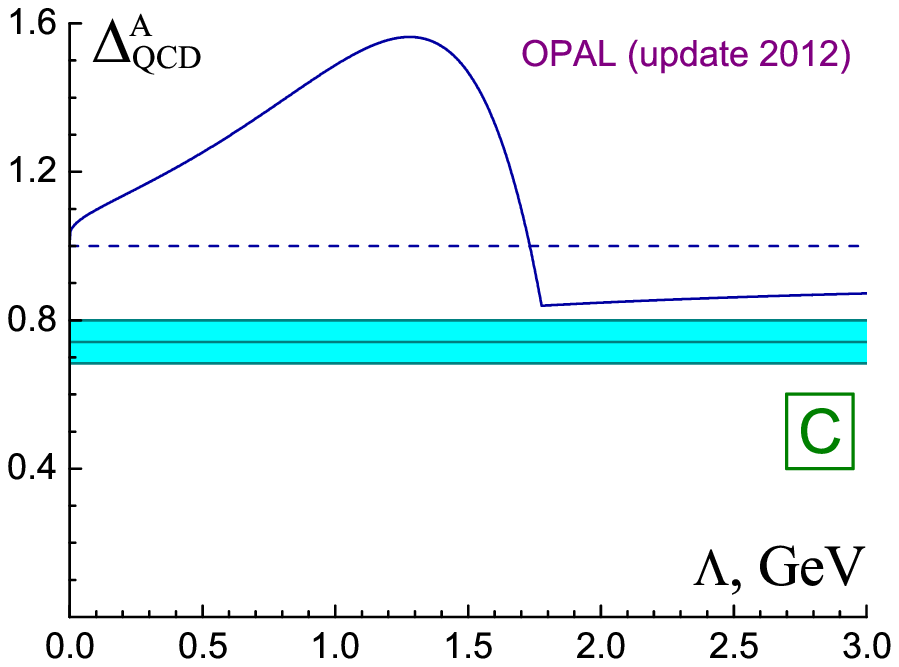}~ &
~\includegraphics[width=72.5mm]{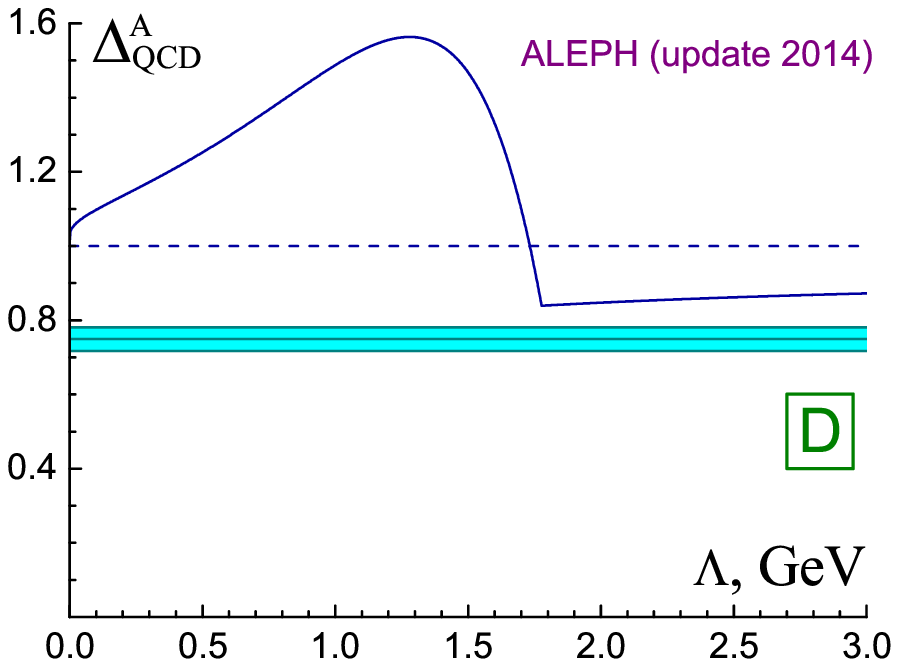}
\end{tabular}
\caption{Comparison of the perturbative expression~$\Delta\ind{\mbox{\tiny
V/A}}{pert}$~(\ref{DeltaQCDPert}) (solid curves) with relevant experimental
data (horizontal shaded bands). Vertical dashed bands denote solutions for the
QCD scale parameter~$\Lambda$. The plots A,~C and B,~D correspond to
experimental data~\cite{OPAL9912} and~\cite{ALEPH0514}, respectively.}
\label{Plot:RTauPT}
\end{figure}

It is worthwhile to mention that the perturbative description of the
inclusive $\tau$~lepton hadronic decay completely leaves out the effects
due to the nonvanishing hadronic production threshold. Moreover, the
perturbative approach suffers from its inherent difficulties, such as the
infrared unphysical singularities. These facts eventually result in the
identity of the perturbative predictions for functions~(\ref{DeltaQCDDef})
in vector and axial--vector channels (i.e.,~$\Delta\ind{\mbox{\tiny
V}}{pert} \equiv \Delta\ind{\mbox{\tiny A}}{pert}$), that contradicts
experimental data. In particular, within perturbative approach the
expression~(\ref{DeltaQCDDef}) acquires the form (in~what follows the
one--loop level with $\nf=3$~active flavors is assumed) \vskip-1mm
\begin{equation}
\label{DeltaQCDPert}
\Delta\ind{\mbox{\tiny V/A}}{pert} = 1 + \frac{4}{\beta_{0}}\!\int_{0}^{\pi}
\frac{\lambda A_{1}(\theta)+\theta A_{2}(\theta)}{\pi(\lambda^2+\theta^2)}
\,d\theta,
\end{equation}
where $\lambda = \ln \bigl( \MTau^2/\Lambda^2 \bigr)$, and
\begin{equation}
A_{1}(\theta) = 1 + 2\cos(\theta) - 2\cos(3\theta) - \cos(4\theta),
\qquad
A_{2}(\theta) = 2\sin(\theta) - 2\sin(3\theta) - \sin(4\theta),
\end{equation}
see Refs.~\cite{PRD88, QCD14}. Furthermore, the perturbative approach is
incapable of describing the experimental data on the inclusive semileptonic
branching ratio in axial--vector channel, see Fig.~\ref{Plot:RTauPT} and
Table~\ref{Tab:RTau}. It is worth noting also that for vector channel
perturbative approach returns two equally justified solutions for the QCD scale
parameter~$\Lambda$, one of which is commonly discarded, see paper~\cite{PRD88}
and references therein for the details.

\begin{figure}[t]
\begin{tabular}{cc}
\includegraphics[width=72.5mm]{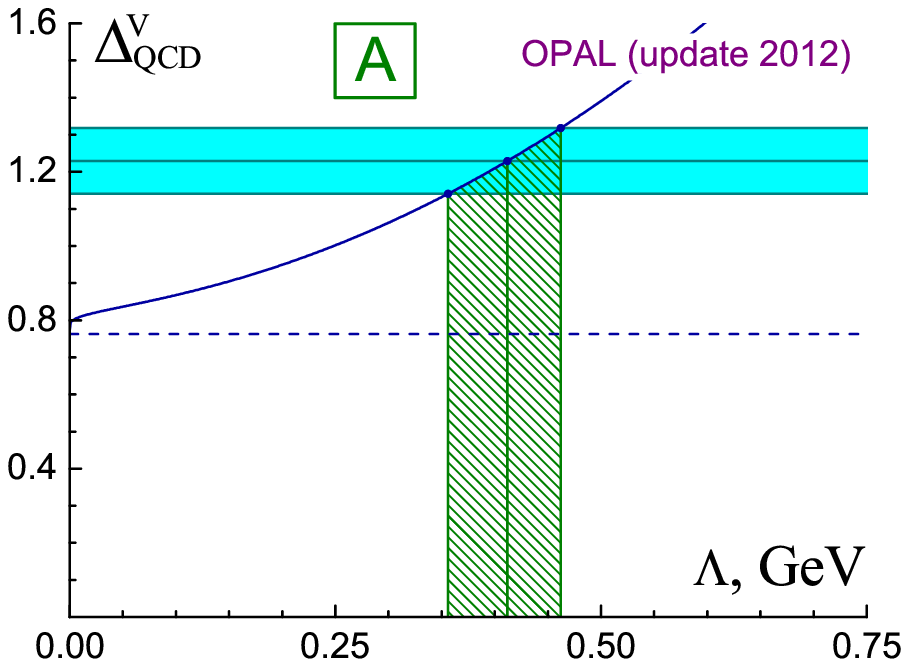}~ &
~\includegraphics[width=72.5mm]{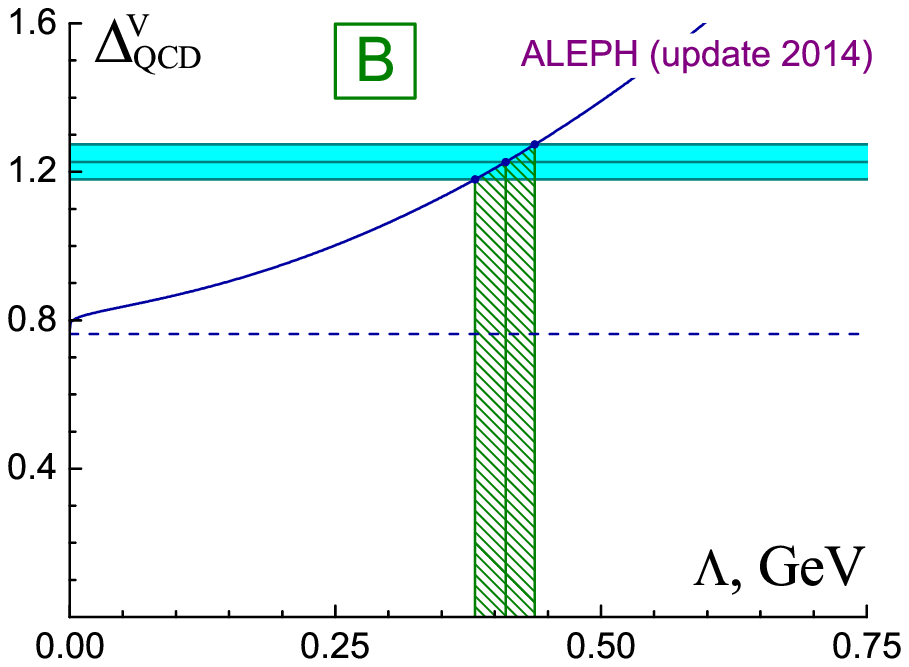} \\[5mm]
\includegraphics[width=72.5mm]{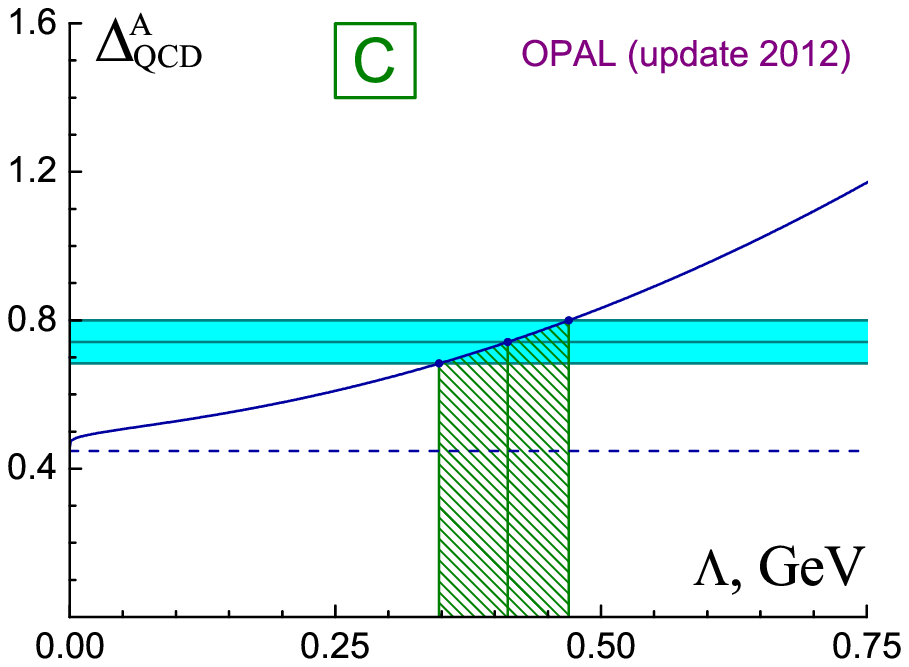}~ &
~\includegraphics[width=72.5mm]{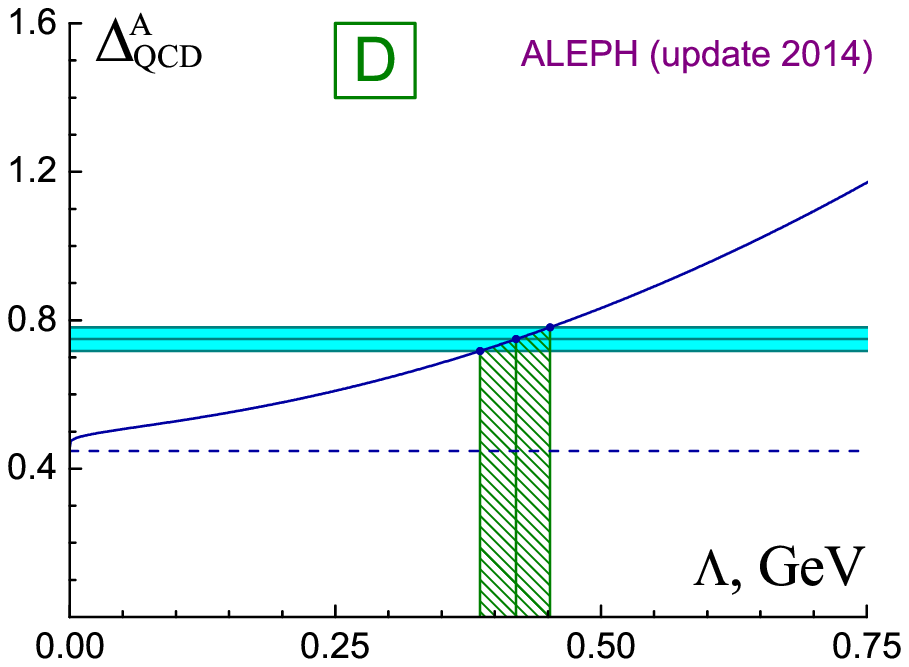}
\end{tabular}
\caption{Comparison of the expression~$\DeltaQCD{V/A}$~(\ref{DeltaQCD})
(solid curves) with relevant experimental data (horizontal shaded bands).
Vertical dashed bands denote solutions for the QCD scale
parameter~$\Lambda$. The plots A,~C and B,~D correspond to experimental
data~\cite{OPAL9912} and~\cite{ALEPH0514}, respectively.}
\label{Plot:RTauDQCD}
\end{figure}

The inclusive $\tau$~lepton hadronic decay was also studied within analytic
perturbation theory and a number of its modifications~\cite{MSS, TauAPT1,
TauAPT2}. However, these papers basically deal either with the total sum of
vector and axial--vector terms~(\ref{RTauGen}) or with the vector term only.
Additionally, APT disregards valuable effects due to nonvanishing hadronic
production threshold and, similarly to perturbative approach, yields identical
predictions for functions~(\ref{DeltaQCDDef}) in vector and axial--vector
channels. For the vector channel APT returns a rather large value for the QCD
scale parameter~($\Lambda \simeq 900\,$MeV). As~for the axial--vector channel,
the APT fails to describe the experimental data on the inclusive $\tau$~lepton
hadronic decay, since for any value of~$\Lambda$ the APT expression for
function~(\ref{DeltaQCDDef}) exceeds its experimental measurement, see also
Ref.~\cite{DQCD3}.

\begin{table}[t]
\centerline{\begin{tabular}{ccccc}
\hline
&
\multicolumn{2}{c}{Perturbative approach\rule[-4pt]{0pt}{4pt}}
&
\multicolumn{2}{c}{Dispersive approach}
\\
& OPAL~\cite{OPAL9912}
& ALEPH~\cite{ALEPH0514}
& OPAL~\cite{OPAL9912}
& ALEPH~\cite{ALEPH0514} \\[-1.75mm]
& {\tiny (update~2012)}\rule{0pt}{9pt}
& {\tiny (update~2014)}
& {\tiny (update~2012)}
& {\tiny (update~2014)}
\\ \hline
\centering
Vector channel\rule{0pt}{9pt}
& $445_{-230}^{+201}$
& $439_{-119}^{+110}$
& $409 \pm 53$
& $409 \pm 28$
\\[1mm]
\centering
Axial--vector channel\rule[-3pt]{0pt}{3pt}
& \multicolumn{2}{c}{no solution}
& $409 \pm 61$
& $419 \pm 33$
\\ \hline
\end{tabular}}%
\caption{Values of the QCD scale parameter~$\Lambda$~[MeV] obtained within
perturbative and dispersive approaches from OPAL~\cite{OPAL9912} and
ALEPH~\cite{ALEPH0514} experimental data on inclusive $\tau$~lepton hadronic
decay (one--loop level, $\nf=3$~active flavors), see Refs.~\cite{PRD88,
QCD14}.}
\label{Tab:RTau}
\end{table}

The dispersive approach to QCD (contrary to perturbative and analytic
approaches) properly accounts for the effects due to nonvanishing hadronic
production threshold. The hadronic contribution~(\ref{DeltaQCDDef}) to the
inclusive semileptonic branching ratio within dispersive approach can
eventually be represented as
\begin{equation}
\label{DeltaQCD}
\DeltaQCD{V/A} = 3\,g_{1}\!\biggl(\frac{\cva}{2}\biggr)\sqrt{1-\cva}
- 3\,g_{2}\!\biggl(\frac{\cva}{4}\biggr)
\ln\biggl(\sqrt{\cva^{-1}}+\sqrt{\cva^{-1}-1}\biggr)
+ \int_{m\va^{2}}^{\infty}\!G\Bigl(\frac{\sigma}{M_{\tau}^{2}}\Bigr)\,
\rho(\sigma)\,\frac{d \sigma}{\sigma}\,,
\end{equation}
where
$G(x) = g(x)\,\theta(1-x) + g(1)\,\theta(x-1) - g(\chi\va)$,
$g(x) = x (2 - 2x^2 + x^3)$,
$\chi\va = m\va^{2}/\MTau^{2}$, $m_{\mbox{\tiny V}}^{2} \simeq
0.075\,\mbox{GeV}^2$, $m_{\mbox{\tiny A}}^{2} \simeq 0.288\,\mbox{GeV}^2$,
spectral density~$\rho(\sigma)$ is specified in Eq.~(\ref{RhoMod}), and
\begin{equation}
g_{1}(x) = \frac{1}{3} + 4x -\frac{5}{6}x^2 + \frac{1}{2}x^3,
\qquad
g_{2}(x) = 8x(1 + 2x^2 - 2x^3),
\end{equation}
see papers~\cite{PRD88, DQCD3, C12, QCD14} and references therein. The
comparison of Eq.~(\ref{DeltaQCD}) with OPAL (update~2012,
Ref.~\cite{OPAL9912}) and ALEPH (update~2014, Ref.~\cite{ALEPH0514})
experimental data is presented in Fig.~\ref{Plot:RTauDQCD} and the
respective values of the QCD scale parameter~$\Lambda$ are given in
Table~\ref{Tab:RTau}. As one may infer from Fig.~\ref{Plot:RTauDQCD}, the
dispersive approach is capable of describing the experimental
data~\cite{OPAL9912, ALEPH0514} on inclusive $\tau$~lepton hadronic decay
in vector and axial--vector channels. The obtained values of the QCD scale
parameter~$\Lambda$ appear to be nearly identical in both channels, that
testifies to the self--consistency of the approach on hand.

\smallskip

The author is grateful to D.~Boito, R.~Kaminski, B.~Malaescu, E.~Passemar,
M.~Passera, J.~Portoles, and H.~Wittig for the stimulating discussions and
useful comments.

\bibliographystyle{aipproc}

\end{document}